\documentclass{ws-ijmpa}
\usepackage[super,compress]{cite}
\usepackage{hyperref}
\hypersetup{colorlinks,urlcolor=black,citecolor=black,linkcolor=black,filecolor=black}
\usepackage{url}
\usepackage{graphicx}
\usepackage{shorthands}
\usepackage{braket}

\begin{document}
\markboth{Bastian B. Brandt, Marco Meineri}{Effective string description of
confining flux tubes}

%%%%%%%%%%%%%%%%%%%%% Publisher's Area please ignore %%%%%%%%%%%%%%%
%
\catchline{}{}{}{}{}
%
%%%%%%%%%%%%%%%%%%%%%%%%%%%%%%%%%%%%%%%%%%%%%%%%%%%%%%%%%%%%%%%%%%%%

\title{Effective string description of confining flux tubes}

\author{Bastian B. Brandt}

\address{Institut f\"ur Theoretisch Physik, Goethe-Universit\"at Frankfurt,
D-60438 Frankfurt am Main \\
and Institut f\"ur Theoretische Physik, Universit\"at Regensburg, D-93040
Regensburg \\
brandt@th.physik.uni-frankfurt.de}

\author{Marco Meineri}

\address{Scuola Normale Superiore, Piazza dei Cavalieri 7 I-56126 Pisa, Italy, 
INFN - sezione di Pisa \\
 and Perimeter Institute for Theoretical Physics - Waterloo, ON N2L 2Y5, Canada.}

\maketitle

\begin{history}
\received{Day Month Year}
\revised{Day Month Year}
\end{history}

\begin{abstract}
We review the current knowledge about the theoretical foundations of
the effective string theory for confining flux tubes and the comparison of the
predictions to pure gauge lattice data. A concise presentation of the effective
string theory is provided, incorporating recent developments. We summarize the
predictions for the spectrum and the profile/width of the flux tube and their
comparison to lattice data. The review closes with a short summary of open
questions for future research.


\end{abstract}

\ccode{11.15.Ha,11.25.Pm,11.30.Cp,12.10.Dm,12.38.Aw,12.40.Nn}

%\tableofcontents

\section{Introduction}

Confinement is one of the most fundamental properties of Quantum Chromodynamics
(QCD). Yet, assured knowledge about its microscopic nature is still lacking
(e.g.~\refcite{Greensite:2003bk}). A possible mechanism to explain quark
confinement is the formation of a tube of strong chromoelectric flux between
quark $q$ and antiquark $\bar{q}$ (for a mesonic boundstate), leading to a
linearly rising potential with the $q\bar{q}$ distance $R$. The emergence of
this picture has been triggered by the observation that the square masses of the
low lying hadrons, when grouped in so called Regge-trajectories, show a linear
increase with the spin. A simplistic model to explain this assumes that mesons
consist of a rotating energy string
(e.g.~\refcite{Greensite:2003bk,Bali:2000gf}). This observation, among others,
led to the formulation of the first string
theories~\cite{Goto:1971ce,Goddard:1973qh}. Contact to QCD can be made if the
region with large field strength density is squeezed into a flux tube with
approximately negligible width. Evidence for this effect can be found
analytically at strong coupling on the
lattice~\cite{Wilson:1974sk,Kogut:1974ag,Isgur:1984bm}. The flux tube dynamics
at large distances would then be governed by a low energy effective string
theory (EST)~\cite{Nambu:1978bd,Luscher:1980fr,Polyakov:1980ca}.

In the dual-superconductor model for confinement (see~\refcite{Baker:1991bc}),
the flux tube appears as the generalization of Abrikosov vortices of type II
superconductors, following the proposal from~\refcite{Nielsen:1973cs}, due to
the condensation of chromomagnetic
monopoles~\cite{Nambu:1974zg,Mandelstam:1974pi,'tHooft:1979uj}.
Vortex and string pictures have distinct features. The vortex picture predicts
an exponential decrease of the field strength at the border of its core with a
constant penetration length, while the EST predicts a Gaussian shape and a
logarithmic growths of the width~\cite{Luscher:1980iy} with $R$. Nonetheless,
the reality is possibly somewhere in between, meaning that the flux tube
consists of a solid, vortex-like, inner core, whose long-distance dynamics is
governed by the EST.

Historically, there are two main frameworks for the construction of the EST.
The \emph{static gauge} approach, pioneered by L\"uscher and
Weisz~\cite{Luscher:1980fr,Luscher:1980ac,Luscher:2004ib} (LW), favors an
explicitly unitary description, while the \emph{orthogonal gauge} formalism,
introduced by Polchinski and Strominger~\cite{Polchinski:1991ax}
(PS), leads to a more symmetric action, but introduces non physical degrees of
freedom. The relation between the two has been recently
elucidated.\cite{Dubovsky:2012sh,Aharony:2013ipa}
The properties of long flux tubes can be investigated in lattice simulations of
pure gauge theories, where the absence of dynamical quarks prohibits string
breaking. The continuum limit (lattice spacing $a\to0$) corresponds to the
weak coupling limit. In going from strong to weak coupling, the theory passes
through a \emph{roughening} phase transition, across which the continuum
symmetries are effectively restored.\cite{Luscher:1980ac}
Simulations have to be done in the continuum phase. Since the first simulations
early in the
80`s~\cite{Lang:1982tj,Griffiths:1983ah,Stack:1983cw,Campbell:1984rm,
Campbell:1984fe,deForcrand:1984wzs}, they have reached a precision which allows
to resolve non-universal terms in the EST. Note, however, that the potential in
the real world will be affected by the presence of dynamical quarks.

The review is organized as follows: In the next section we will discuss the
construction of the EST, with particular emphasis on recent developments.
Section~\ref{sec:predict} is devoted to the EST predictions concerning the
spectrum and the width of the flux tube and in section~\ref{sec:test} we discuss
the comparison to lattice data. We will conclude with a summary and discussion,
where we try to identify relevant open questions and ideas for future research.
Limitations in space urged us to subjectively pick topics from a vast field. We
would like to apologize to all those whose contribution has not been appreciated
sufficiently. A small remark on related topics is included after the
conclusions.
\vspace{-0.15cm}

\section{The effective action for a long confining string}
\label{sec:action}

Our aim is to write down the low energy effective field theory (EFT) for stable
non-interacting confining flux tubes. Two possible realizations are a flux tube
stretched between two static sources or a closed flux tube wrapping around a
compactified dimension, rendering it stable under contraction. The latter does
not appear in real world QCD, but can be realized at finite volume on the
lattice. Since a long flux tube is approximately a one-dimensional object, the
EFT will be a two dimensional theory, taking the usual form of a derivative
expansion. The degrees of freedom are the Goldstone bosons (GB) associated with
the breaking of the translational symmetry by the tube, i.e. the quantized
transverse oscillation modes. They carry a minimal energy of $O(R^{-1})$, so
that the EFT is accurate for long flux tubes.

The EST is expected to break down at some scale of order $\La_\textup{QCD}$
(i.e. $\sqrt{\si}R \gtrsim 1$ -- see below). String breaking and (virtual)
glueball emission are not accounted for. Both processes are suppressed at large
$N$, and on-shell glueball emission also at low energies. More massless
particles may arise if additional symmetries are broken by the string, as for
instance supersymmetry. Weakly coupled light massive modes, if present, can be
added to the action as well (see sec.~\ref{subsec:th:spectrum}). Sometimes
the effective action can be derived explicitly, by matching to a weakly
coupled UV completion,\cite{Akhmedov:1995mw} or in cases in which a weakly
curved holographic background is available.\cite{Aharony:2009gg,Aharony:2010cx}
\vspace{-0.1cm}

\subsection{Lorentz symmetry and the effective string action.}
\label{subsec:static}

We consider a flux tube which extends in the $(x^0,x^1)$ plane. We use indices
$\mu,\nu,\ldots$ for the full $D$-dimensional space, indices
$\alpha,\beta,\ldots$ for the $(0,1)$ plane and latin ones $i,j,\ldots$ for the
$D-2$ transverse directions. The action for the GB, which we denote $X^i$,
retains an $ISO(1,1)\times SO(D-2)$ symmetry group, combining Poincar\'e
symmetry on the worldsheet and rotational symmetry in the transverse directions.
Under transverse translations $X^i \to X^i+\epsilon^i$, so the GB couple
derivatively. When constructing the action, one can omit terms proportional to
lower order equations of motion (EOM), which can be swept to higher orders by
field redefinitions. Thus, for a worldsheet $\mathcal{M}$ the first few terms in
the action are
\begin{multline}
S_\textup{bulk}  = \int_\mathcal{M}\!\!
 \displaystyle \big[ -\si -\frac{\si}{2}\,\pa_\al X^i \pa^\al X^i + c_2
(\pa_\al X^i
\pa^\al X^i)^2 + c_3 (\pa_\al X^i \pa_\be X^i)^2\\
  \: \: \: \displaystyle +c_4 (\pa_\al\pa_\be X^i \pa^\al \pa^\be X^i)
(\pa_\ga X^j\pa^\ga X^j)\dots \big] .
\label{sbulkfew}
\end{multline}
Here $\si$ is the string tension, governing the linear rise of the potential,
which for SU(3) in $D=4$ assumes a value of
$\si^{-1/2}\approx0.42$~fm~\cite{Necco:2001xg}. This value is used to convert to
\emph{physical} units in the following, even when we are in an unphysical
situation. Open flux tubes end on static sources, leading to an additional
boundary action, constrained by Dirichlet boundary conditions ($\pa_\parallel
X^i=0$). The first terms are
\begin{eqnarray}
S_\textup{boundary} =\! \displaystyle \int_{\pa \mathcal{M}}\!  \big[ \m + b_1
\pa_\perp X^i \pa_\perp X^i\!
+b_2 \pa_\parallel\pa_\perp X^i \pa_\parallel\pa_\perp X^i + b_3(\pa_\perp
X^i\pa_\perp X^i)^2\dots \big]
\label{sboundfew}
\end{eqnarray}
The string also breaks Lorentz transformations in mixed directions $(\alpha,i)$,
which must be nonlinearly realized as well (no additional GB are
required\cite{Low:2001bw}). The consequent constraints on the coefficients $b_i$
and $c_i$ were first investigated via open-closed duality\cite{Luscher:2004ib}
(see also \refcite{Cohn:1992nu}). While this method only exploits $SO(D-2,1)$
invariance,\cite{Meyer:2006qx,Aharony:2010cx}\footnote{Indeed, the crucial
ingredient is a relativistic dispersion relation for the closed string states,
$E^2\!=\!E(P_\perp\!\!=\!0)^2\!+\!P_\perp^2$. $P_\perp$ is the transverse
momentum.\label{disprel}} Lorentz invariance can be implemented directly in the
action, in line with the standard EFT
construction.\cite{Coleman:1969sm,Callan:1969sn,Volkov:1973vd}
One constructs fields which, under $ISO(D-1,1)$, only change via a
field (i.e. also coordinate) dependent element of the unbroken
subgroup.\cite{Coleman:1969sm} These are the induced metric on the worldsheet
$g_{\al\be}=\pa_\al X^\m\pa_\be X_\m$ and the extrinsic curvature
$K_{\al\be}^I=n^I_\m\nabla_\al\pa_\be X^\m$, where  $X^\m=(x^\al,X^i)$ and
$n^\m_I$ is a $(D-2)$-tuple of normal vectors. A Lorentz transformation in the
plane $(\al,i)$ acts as a diffeomorphism on the worldsheet coordinate, $\de
x^\al = \ep^\al{}_i X^i(x)$, so the bulk effective action is diffeomorphism
invariant,\footnote{Another systematic way of obtaining $S_\textup{bulk}$
employs directly the nonlinear transformation rule of the physical dofs
$X'^i(x')= X^i(x)+\ep^i{}_\al\, x^\al$. Plugging $\de x^\al=\ep^\al{}_i X^i(x)$
back in, one gets the in-form variation $\de X^i=\ep^{j\al} (\de_j^i
x_\al+\pa_\al X^i X_j(x))$. Once applied to a term invariant under
$ISO(1,1)\times SO(D-2)$ like those in eq. \eqref{sbulkfew}, a recurrence
relation is
generated.\cite{Aharony:2010cx,Aharony:2011gb,Gliozzi:2012cx,Cooper:2013kga}
See also \refcite{Gomis:2012ki}.} i.e. 
\beq
S_\textup{bulk} = 
-\int_\mathcal{M} \sqrt{-\textup{det} g} \left(\si+\al\, K_{\al\be}^I
K^{\al\be}_I +\dots\right).
\label{sbulkdiff}
\eeq
Contrary to eq. \eqref{sbulkfew}, worldsheet indices are here contracted through
$g^{\al\be}$. The boundary action can be constrained
similarly,\cite{Luscher:2004ib,Billo:2012da}
leading to $b_1=b_3=0$. The first addend in $S_\textup{bulk}$ is the Nambu-Goto
(NG) action,\cite{Goto:1971ce,Nambu:1978bd} whose expansion fixes the $c_i$ in
eq. \eqref{sbulkfew}. In particular, the $c_4$ term vanishes at the level of
this classical analysis. $(K_\al{}^{\al I})^2$ does not appear in
eq.~\eqref{sbulkdiff}, because Gauss-Codazzi equations express it in terms of
$K_{\al\be}^I K^{\al\be}_I$ and the Ricci scalar (which is topological in 2D).
The first non-trivial term in $K_{\al\be}^I K^{\al\be}_I$ up to the free EOM has
eight derivatives, so that corrections to the NG action are strongly suppressed
in the derivative expansion.

One might argue that the action~\eqref{sbulkdiff} directly follows from the
freedom of choosing coordinates on the worldsheet. However, the gauge fixing
procedure often leads to pathologies away from the critical dimension $D=26$ (or
$D=3$\cite{Mezincescu:2010yp}). For instance, the gauge freedom of the NG action
can be fixed completely, leading to a quadratic
action\cite{Goddard:1973qh,Arvis:1983fp} and to the
light-cone\cite{Polchinski:1998rq} (LC) spectrum \eqref{ELC}, but this choice
breaks the symmetry down to $ISO(1,1)\times SO(D-2)$. The EFT construction
expresses the action \eqref{sbulkdiff} in terms of the physical degrees of
freedom (DOF), through the \emph{static gauge} choice $X^\m=(x^\al,X^i)$, and
requires no gauge fixing. The theory can be regularized in a Lorentz invariant
way in any number of dimensions via dimensional
regularization.\cite{Dubovsky:2012sh} If, instead, the regularization breaks
Lorentz symmetry, finite counterterms must be included to obtain, say, the
correct spectrum \eqref{ESTspectrum} of the Lorentz invariant theory. The
leading one turns out to be proportional to the $c_4$ term in
eq.~\eqref{sbulkfew}.\cite{Aharony:2011gb,Dubovsky:2012sh,Aharony:2013ipa}
Zeta function regularization, associated with Weyl ordering, requires
$c_4=\si/8\pi $~\cite{Dubovsky:2012sh} while any continuum regulator which
preserves the number of worldsheet dimensions leads instead to
$c_4=\si(D-26)/192\pi$.\cite{Aharony:2013ipa}.
\vspace{-0.15cm}

\subsection{The action in orthogonal gauge and the critical dimension}
\label{subsec:PS}

Consistent quantization of the action can also be achieved in orthogonal
gauge, i.e. $g_{++}=g_{--}=0$ in light-cone coordinates. The gauge fixed form of
the NG action, $S_\textup{NG}= -2\si \int_\mc{M} \pa_+ X^\m \pa_- X_\m$, is
explicitly Lorentz invariant and conformal (a remnant of diffeomorphism
invariance) but, in principle, receives gauge-fixing contributions from a $bc$
ghost system and the path-integral measure. In~\refcite{Polchinski:1991ax} PS
found those contributions directly, by adding counterterms which render the
covariant quantization of $S_\textup{NG}$ consistent, i.e. leading to a central
charge $c=26$. The additional terms can be non-polynomial, as long as they are
local in a long string expansion, which in this gauge demands that $\pa_\pm
X^\m=O(R)$. The lowest order modification of the free action, up to the free EOM
and terms proportional to the
constraints is\cite{Polchinski:1991ax,HariDass:2007dpl}
\beq
S_\textup{bulk}^\textup{PS} =  \int_\mc{M}-2\si \pa_+ X^\m \pa_- X_\m+
\frac{26-D}{48\pi}\frac{\pa_+\pa_- (\pa_+X^\m \pa_- X_\m)}{\pa_+X^\m \pa_-
X_\m}+\ldots
\label{PSaction}
\eeq
The coefficient is fixed by requesting $c=26$. The PS action shows the special
role of $D=26$: in this case the theory has a chance of being UV complete, while
in general it breaks down for short strings. Since the derivation of the action
\eqref{PSaction} is heuristic (but for specific examples\cite{Natsuume:1992ky}),
equivalence with the static gauge action must be checked. In theories with
holographic duals the gauge fixing can be explicitly done in the bulk, and the
equivalence on the boundary follows.\cite{Aharony:2013ipa} A general conclusion
can be reached by computing gauge invariant observables in the two formalisms.
This comparison was carried out in \refcite{Dubovsky:2012sh}, by computing the
$2\to2$ scattering of transverse modes. In particular, the PS interaction term
is generated as a finite part of the amplitude in static gauge, thus showing
that to one loop order the 1PI actions of the two theories agree. Incidentally,
it was also noticed that the PS term is the leading order one responsible for
annihilation. This will be important in subsec. \ref{subsec:th:spectrum}.

Let us finally mention that the PS treatment of non-critical strings is
reminiscent of the one by Polyakov,\cite{Polyakov:1981rd} in which a Liouville
mode arises, away from $D=26$, from the auxiliary worldsheet metric
$\ga_{\al\be}$, which cannot be gauged away due to the Weyl anomaly. The PS
action is simply obtained by identifying $\ga_{+-}=g_{+-}$. In
\refcite{Hellerman:2014cba}, the authors build on this relation, and systematize
the subleading corrections to eq.~\eqref{PSaction}.

\section{Predictions from the effective action}
\label{sec:predict}

The effective string action~\eqref{sbulkdiff} can be used to compute low energy
observables and compare the results with lattice data. In pure gauge theories, a
static $q\bar{q}$ pair is represented by two Wilson lines winding around the
temporal lattice, a Polyakov loop correlation function. Polyakov loop
correlators can also wind around a spatial dimension, in which case they
represent the creation (or annihilation) operator for a closed flux tube, whose
temporal correlator can be used to investigate closed strings states. For the
investigation of open string excited states it is often beneficial to use Wilson
loops, including creation and annihilation operators for flux tubes with given
quantum numbers. While even the partition functions of such observables are
suitable for the investigation of the EST, we will concentrate on the flux tube
spectra, which contain similar information and are of immediate physical
interest. We also comment on the transversal shape and the width of the string.
The shape can be computed by a correlation function of a flux tube state with
components of the energy momentum tensor, represented by a single plaquette, for
instance.

\subsection{Spectrum of the flux tube.}
\label{subsec:th:spectrum}

Perturbatively, states are labeled by the number $n_m$ and $\bar{n}_m$ of free
left and right moving phonons with wave number $m$, each phonon carrying an
index $i$ associated with the transverse direction of oscillation.\footnote{In
fact, in orthogonal gauge physical states are transverse only at leading order
in $R$. There is however a one-to-one correspondence with transverse
states.\cite{Aharony:2011ga}} It is useful to define the level $N=\sum_m m n_m$
and similarly for $\bar{N}$. The states are organized in irreducible
representations of the group $SO(D-2)$ of transverse rotations. Charge
conjugation\footnote{To be precise: $C$ also includes a reflection of the
chromomagnetic flux.} ($C$) exchanges $N$ and $\bar{N}$, while we refer as
parity ($P$) to the inversion of the transverse coordinates. For open strings
$N=\bar{N}$, since the longitudinal momentum $q=N-\bar{N}$ vanishes. We ignore
the transverse momentum for closed strings (see footnote \ref{disprel}). Both,
closed and open string spectra have been computed in static
gauge~\cite{Luscher:2004ib,Aharony:2010db}, while the analysis in orthogonal
gauge is restricted to closed
strings~\cite{Polchinski:1991ax,Drummond:2004yp,HariDass:2006sd,Aharony:2011ga}.
Recently, a new approach based on the Thermodynamic Bethe Ansatz (TBA) has been
put
forward.\cite{Dubovsky:2012wk,Dubovsky:2013gi,Dubovsky:2014fma,Caselle:2013dra}
An observation related to the TBA approach provides a fast way to compute the
spectra up to $O(R^{-5})$. The PS annihilation term mentioned in subsec.
\ref{subsec:PS} is the first deviation of the infinite volume scattering from a
factorisable S-matrix,\cite{Dubovsky:2012wk} \footnote{The interplay between
integrability and Lorentz invariance has also lead to the search for theories,
in which additional particles restore the former without breaking the
latter.\cite{Dubovsky:2015zey}} which yields, via TBA, the light-cone spectrum
mentioned in subsec. \ref{subsec:static},
\begin{equation}
E^\textup{LC}_{\textup{closed}\atop\textup{open}} = \sqrt{(\si
R)^2+\kappa_{\textup{closed}\atop\textup{open}}\, \si
\left(\frac{N+\bar{N}}{2}-\frac{D-2}{24}\right)+
{\biggr(\frac{2\pi q}{R}\biggr)}^2},
\label{ELC}
\end{equation}
with $\kappa_\textup{closed}=8\pi$ and $\kappa_\textup{open}=2\pi$. The open LC
spectrum is obtained by supplementing the S-matrix with the simplest consistent
boundary reflection factor.\cite{Caselle:2013dra} In static gauge, by canceling
the leading order PS amplitude with a tree level counterterm, one finds the
action yielding the LC spectrum up to six
derivatives:\cite{Dubovsky:2012sh,Dubovsky:2014fma}
\beq
S_\textup{LC} = S_\textup{NG} + \frac{D-26}{192\pi}\int_\mathcal{M}
\pa_\al\pa_\be X^i \pa^\al \pa^\be X^i \pa_\ga X^j
\pa^\ga X^j + \dots 
\label{PSstatic}
\eeq
In this equation, $S_\textup{NG}$ is the NG action in static gauge, \emph{plus}
all the counterterms necessary to regulate it in a Lorentz invariant way (see
the end of section~\ref{subsec:static}). In static gauge, terms with $k$
derivatives contribute from order $R^{-k+1}$, so that the closed string spectrum
is expected to deviate from the LC one at $O(R^{-5})$. For open strings, the
contribution proportional to $b_2$ in~\eqref{sboundfew} can either be computed
by diagonalizing the Hamiltonian~\cite{Aharony:2011ga} or via
TBA.\cite{Caselle:2013dra} Consequently, the spectrum up to $O(R^{-5})$ is
\beq
E_{\textup{closed}\atop\textup{open}} =
E^\textup{LC}_{\textup{closed}\atop\textup{open}}
+b_2\frac{\pi^3}{R^4} \mathcal{E}^{b_2}_{\textup{closed}\atop\textup{open}}
-\frac{\pi^3(D-26)}{48 \si^2R^5}\,
\mathcal{E}^\textup{PS}_{\textup{closed}\atop\textup{open}}
+O(R^{-7}).
\label{ESTspectrum}
\eeq
Both the $b_2$ term and the PS interaction lift the degeneracies between
different irreducible representations of $SO(D-2)$. The corrections have been
computed for the first few levels.\cite{Aharony:2010db,Aharony:2011ga}
$\mathcal{E}^\textup{PS}$ vanishes for the ground-state of both open and closed
strings, as well as for closed string excited states which do not contain both
left and right movers, and, in general, for 3D. For closed and open strings, the
lowest excited states which are affected are:\footnote{Two contributions at
$O(R^{-7})$ were computed as well for closed strings,\cite{Aharony:2010db} but
the spectrum is not complete at this order.}
\begin{gather}
\mathcal{E}^{b_2}_\textup{closed}=0, \quad
\mathcal{E}^\textup{PS}_\textup{closed} = -64 \times
\begin{system}
D-3 \\
-1 \\
1
\end{system}\quad
\begin{array}{l}
\textup{scalar}, \\
\textup{symm. traceless}, \\
\textup{a.-symm.}
\end{array}
\quad (N,\bar{N})=(1,1)\,,
\label{epPSclosed} \\
%\eeq
%\beq
\mathcal{E}^{b_2}_\textup{open} =
\begin{system}
- \frac{D-2}{60} \\
- \frac{D-2}{60}-4 \\
- \frac{D-2}{60}-8 \\
 - \frac{D-2}{60}-32 \\
 - \frac{D-2}{60}-8
\end{system},\quad
\mathcal{E}^\textup{PS}_\textup{open} =
\begin{system}
0 \\
D-3  \\
0 \\
16 (D-3) \\
4 (D-2)
\end{system}\quad
\begin{array}{l}
N=0, \\
N=1, \ \textup{vector} \\
N=2, \ \textup{scalar} \\
N=2, \ \textup{vector} \\
N=2, \ \textup{symm. traceless}\,.
\end{array}
\label{epopen}
\end{gather}

The TBA procedure of extracting the spectrum from the scattering amplitudes also
allows the inclusion of possible light massive modes. One possibility (suggested
by lattice measurements, see figure~\ref{fig:b2_plot}) is a light $CP=--$
pseudoscalar state,\cite{Dubovsky:2013gi} known as the worldsheet axion because,
as a consequence of Lorentz invariance, it couples to the self-intersection
number.\cite{Polyakov:1986cs} Furthermore, the TBA method relies on a
low-momenta expansion which has, per se, better convergence
properties,~\cite{Dubovsky:2014fma} and provides
insight\cite{Dubovsky:2013gi,Dubovsky:2014fma} on a surprising fact. The
comparison to lattice data shows qualitative agreement with the full LC spectrum
for strings as short as $R\sim 1/\sqrt{\si}$, way below the radius of
convergence of its expansion in $R^{-1}$ for the excited states. In particular,
there is no justification for the use of eq.~\eqref{ELC} in the derivative
expansion. In contrast, it is the \emph{full} LC spectrum which is generated
from the leading order in momenta of the S-matrix in the TBA procedure.
Unfortunately, the breaking of integrability at higher orders is likely to make
the TBA machinery complicated.

The squared extrinsic curvature in the action \eqref{sbulkdiff} (the so called
\emph{rigidity} term) starts at four derivatives, and its leading order term
vanishes up to the free EOM. Nevertheless, its contribution to the ground state
was found to be non
vanishing.\cite{Polyakov:1986cs,German:1989vk,Ambjorn:2014rwa,
Caselle:2014eka} The result admits an expansion in $1/\al$, and so there is no
contradiction with the fact that this term is trivial order by order in the long
string expansion (being also an expansion in $\al$). However, compatibility with
the target space Lorentz invariance has not been checked, and the form of the
correction coincides with the one of a massive particle, so that care is needed
in interpreting the results.

\subsection{Width of the flux tube.}
\label{subsec:th:width}

Quantum fluctuations provide the two-dimensional worldsheet with an effective
width, which is an IR effect computable within the EST and must be distinguished
from the ``intrinsic'' UV scale $1/\sqrt{\si}$. The profile of the string is
associated with the expectation value of the chromo-electric energy density
$\mathcal{E}(x)\propto
\bra{q\bar{q}}\textup{Tr}E^2(x)\ket{q\bar{q}}-\bra{0}\textup{Tr}E^2(x)\ket{0}$,
which can also be reconstructed from its moments,
\beq
w^{2n}(x^\al)=\braket{(X^i(x^\al)-\bar{X}^i)^{2n}} \sim 
 \frac{\int d^{D-2} x_\perp\, x_\perp^{2n}\, \mathcal{E}(x)}{\int d^{D-2}
x_\perp \mathcal{E}(x)},\quad \bar{X}^i = \frac{1}{\mathcal{A}} \int_\mathcal{M}
X^i.
\label{wmoments}
\eeq
The width is conventionally defined by the second moment, the variance, measured
at the middle point of the $q\bar{q}$ axis. It is UV divergent, so that
typically a point splitting procedure is applied for its regularization. The
computation in the free theory~\cite{Luscher:1980iy,Caselle:1995fh} predicts a
Gaussian profile and a logarithmic broadening of the width with an universal
coefficient
\beq
w_0^2=\frac{D-2}{2\pi \si} \log \left(R/R_\textup{UV}\right)+\dots \,,
\label{w0}
\eeq
where $R_\textup{UV}$ is a UV scale. Consequently, as mentioned in the
introduction, infinitely long flux tubes are delocalized, which is a
manifestation of the absence of Goldstone bosons in the thermodynamic limit in
2D.\cite{Mermin:1966fe,Coleman:1973ci} The dots in eq.~\eqref{w0} stand for
terms subleading in $R$, which have been computed for toroidal and cylindrical
geometries up to the leading order contributions of the $c_2$ and $c_3$ terms
in~\refcite{Gliozzi:2010zt}. As mentioned earlier, the EST predictions differ
from the prediction of a classical vortex model. In the light of a fusion to a
vortex/string picture the width of the EST may be thought of as an additional
quantum contribution, dominating for long strings.

\section{Comparison between lattice results and the EST}
\label{sec:test}

We will now come to the comparison of the EST predictions for the spectrum of
open and closed flux tubes and measurements of the flux tube profile with
lattice results. A discussion of the lattice methods is beyond the scope of the
review and we only mention important aspects in footnotes.

\subsection{Spectrum of the flux tube}

\subsubsection{Overview of results}
\label{sub:spec:over}

\begin{figure}[t]
 \centering
 \begin{minipage}{.45\textwidth}
 \centering
 \includegraphics[width=.80\textwidth]{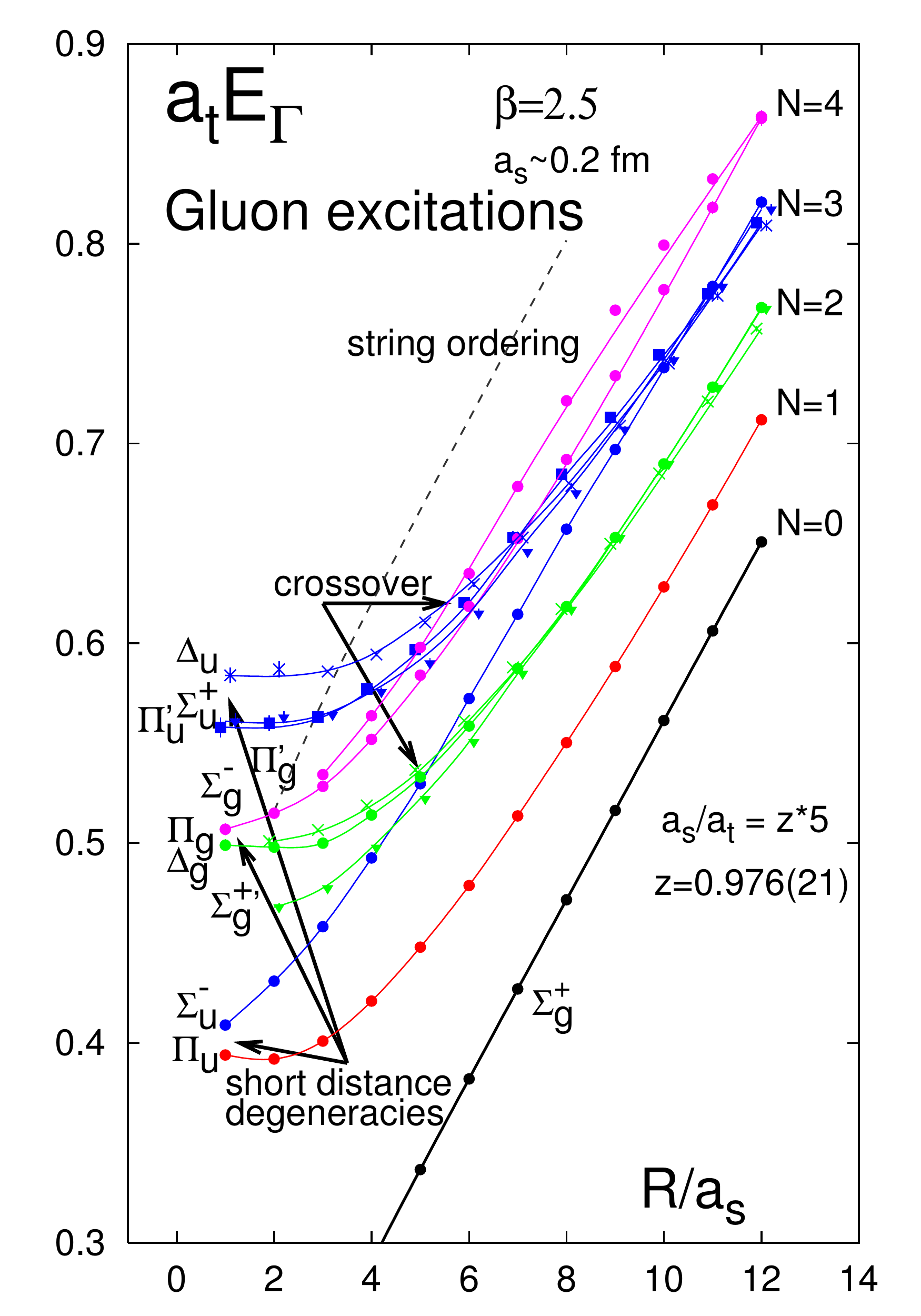}
 \end{minipage}
 \begin{minipage}{.51\textwidth}
 \centering
 \includegraphics[width=.88\textwidth]{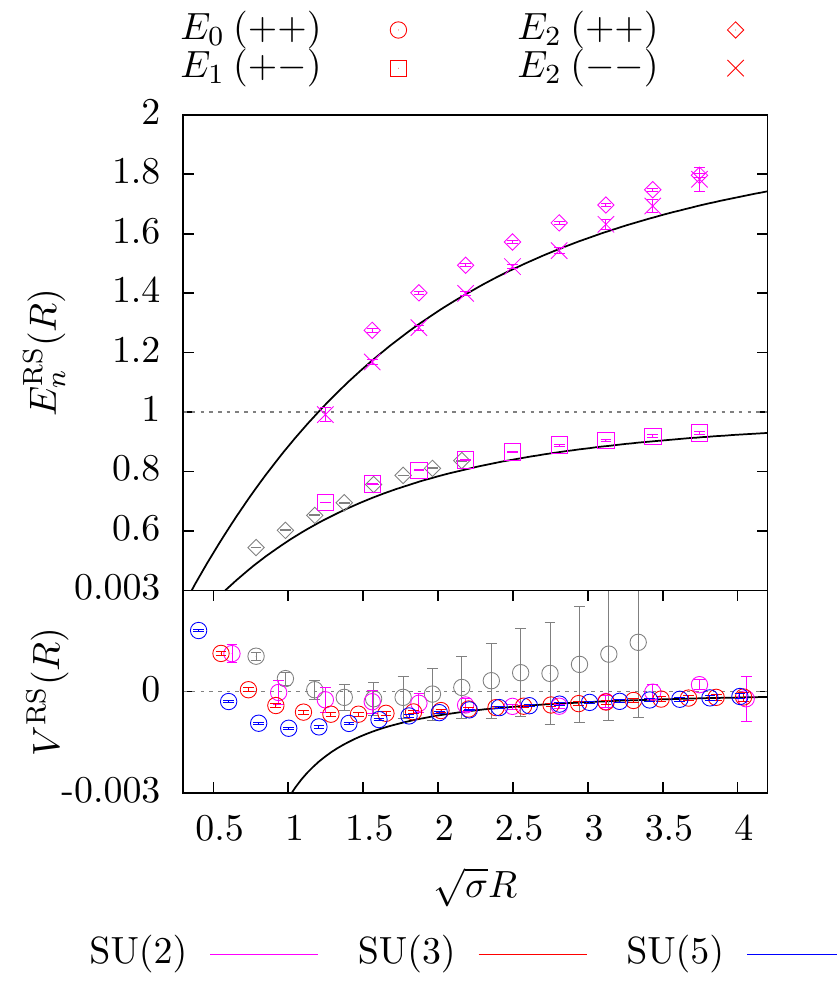}
 \end{minipage}
 \caption{{\bf Right:} Results for the spectrum of the open flux tube in
 4D SU(3) gauge theory from~\citen{Juge:2002br}. States are labeled by the
 projection of angular momentum $J$ on the string axis ($\Sigma,\Pi,\Delta$ for
 $J=0,1,2$), parity $P$ (subscripts $g,u$ for 1,-1) and charge conjugation 
 $C$ (superscript +,-). The study used anisotropic lattices with lattice
 spacings $a_s$ (spatial) and $a_t$ (temporal). {\bf Left:}
 Results for the spectrum of the open flux tube in 3D SU($N=2,3,5$) gauge
 theories from~\citen{Brandt:2010bw,Brandt:2013eua,Brandt:unpublished}
 ($\beta=5.0,\,20.0$ and 54.0, respectively)
 and SU(2) from~\citen{Majumdar:2004qx,HariDass:2007tx} (gray symbols;
 $\beta=7.5$) versus $\sqrt{\sigma}R$. We have
 rescaled the energies such that the leading order expansion of the LC energy
 levels is equal to $n$ (see~\citen{Brandt:2010bw}).
 The states are classified according to $(CP)$. The lower panel shows a
 blow-up of the groundstate energies. The solid lines correspond to the LC
 spectrum.}
 \label{fig:open-enes}
\end{figure}

\begin{figure}[t]
 \centering
 \includegraphics[width=.88\textwidth]{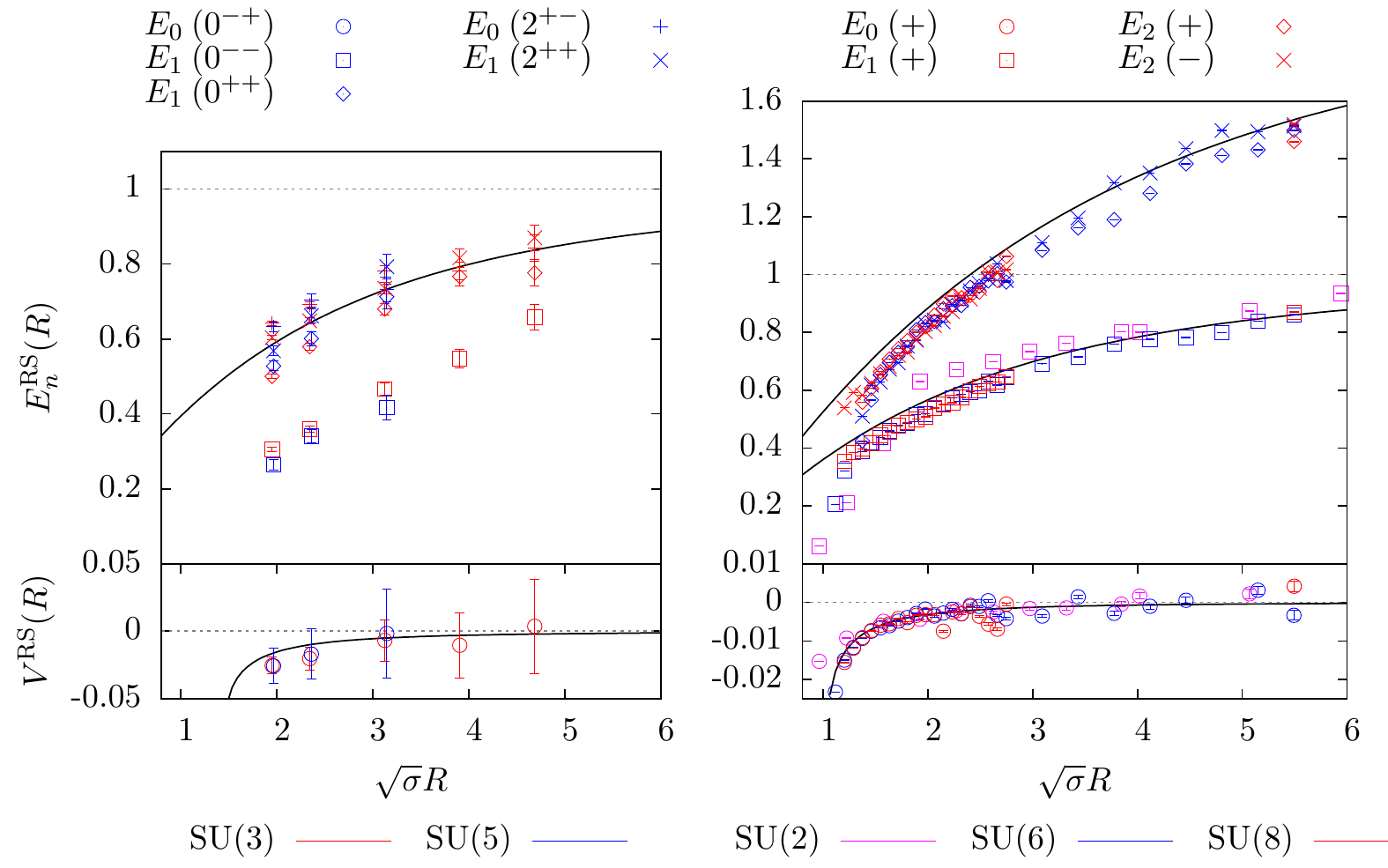}
 \caption{As figure~\ref{fig:open-enes} (right) for the energy levels of the
 closed flux tube in 4D SU($N=3,5$) (left), $\beta=6.0625$ and $17.63$, and 3D
 SU($N=2,6,8$) (right) ($\beta=16,171$ and $306$) gauge theories
 from~\cite{Athenodorou:2010cs,Athenodorou:2016kpd} versus $\sqrt{\sigma}R$. We
 have included states with $q=0$, set
 $n=(N_L+N_R)/2$ and rescaled the energies as in figure~\ref{fig:open-enes}. The
 states in 4D are classified by $(J^{CP})$ and in 3D by $(P)$.}
 \label{fig:closed-enes}
\end{figure}

Since the first studies concerning open flux tubes in the early
80`s~\cite{Lang:1982tj,Griffiths:1983ah,Stack:1983cw,Campbell:1984rm,
Campbell:1984fe,deForcrand:1984wzs} the accuracy and reliability of lattice
measurements~\cite{Otto:1984qr,Hasenfratz:1984gc,Barkai:1984ca,Ambjorn:1984yu,
Sommer:1985du,Huntley:1985ts,Gutbrod:1985be,Itoh:1986gy,Flensburg:1986qc,
Campbell:1987nv,Hoek:1987uy,Ford:1988ki,Perantonis:1988uz,Mawhinney:1989ve,
Perantonis:1990dy,Michael:1990az,Booth:1992bm,Bali:1992ab,Bali:1992ru,
Bali:1994de,Iwasaki:1996sn,Bali:1997am,Beinlich:1997ia,Edwards:1997xf,
Necco:2001xg,Morningstar:1998da,Juge:2002br,Bali:2003jq,Juge:2004xr,
Kuti:2005xg,Lohmayer:2012ue} in 3 and 4D SU($N$) gauge theories has steadily
improved, together with much better control over systematic
uncertainties.\footnote{To make contact with the EST in the continuum there are
certain systematic effects that need to be controlled. A basic effect is the one
due to finite lattice spacing $a$, whose control demands to take the continuum
limit ($a\to0$), which, apparently is rather uncritical. Another effect concerns
the finite extent of the lattice, leading to around-the-world contributions and
finite temperature effects. The most severe effect is the one due to
contaminations from excited states. The contribution of the next excited state
is suppressed by a factor $\exp(-T\Delta E)$, where $T$ is the temporal extent
of the loop and $\Delta E$ the associated energy gap. Since $\Delta E$ decreases
with $R^{-1}$ the problem becomes more severe for large $R$. To reduce
contaminations, there are two approaches: (i) Optimizing the overlap with the
groundstate: Suitable methods to achieve this are
smearing~\cite{Falcioni:1984ei,Hasenfratz:2001hp} and variational or correlation
matrix methods~\cite{Michael:1982gb,Campbell:1984fe,Blossier:2009kd}. (ii) Using
loops with large $T$: Here the signal-to-noise ratio decreases exponentially, so
that powerful error reduction algorithms (see text) are needed.} High accuracy
measurements of large loops have become available by the introduction of the
multilevel algorithm~\cite{Luscher:2001up}, which has since been used
extensively for studies of open flux tube
spectra~\cite{Luscher:2002qv,Majumdar:2002mr,Koma:2003gi,Caselle:2004er,
Majumdar:2004qx,Meyer:2006gm,HariDass:2006pq,HariDass:2007tx,Brandt:2009tc,
Brandt:2010bw,Mykkanen:2012dv,Brandt:2013eua}. Similar studies have also been
performed in 3D U(1)~\cite{Panero:2004zq,Caselle:2014eka} and 3D
Z$_2$~\cite{Caselle:1996ii,Caselle:2002ah,Caselle:2004jq,Caselle:2005vq,
Caselle:2010pf, Billo:2011fd, Billo:2012da} gauge theories. Measurement of the
energy levels of closed flux tubes have started only little
later~\cite{Michael:1986cj,Michael:1988be,Michael:1989vh,Teper:1993gm,
Michael:1994ej,Teper:1998te,Lucini:2001ej,Lucini:2001nv,
Lucini:2002wg,Juge:2003vw,Athenodorou:2007du,Athenodorou:2010cs,
Athenodorou:2011rx,Athenodorou:2016kpd}. We show a collection of results for the
lowest energy levels of open and closed flux tubes in
figures~\ref{fig:open-enes} and~\ref{fig:closed-enes}, respectively.

To briefly summarize the main findings: For (most of) the energy levels the data
shows a remarkable agreement with the full LC predictions down to small values
of $\sqrt{\si}R\sim1$, rather than with its expansion in powers of $R^{-1}$,
independently of the number of dimensions and the gauge
group.\cite{Dubovsky:2014fma} Typically, agreement with the string states is
seen starting from $R\approx0.5$~fm for the groundstate up to 2~fm for the
excited states, increasing with the excitation level. The convergence to the LC
spectrum appears to be enhanced when going to larger values of $N$. These
observations are in full agreement with the discussion of the EST in
sec.~\ref{sec:action} and \ref{sec:predict} and are basically independent of the
lattice spacing and thus should hold in the continuum. With ever increasing
accuracy, deviations from the LC spectrum become visible, once more in agreement
with the EST. Some of those deviations in 4D SU($N$) gauge theories, however,
are not in agreement with the the spectrum in eq. \eqref{ESTspectrum} and could
be signs for the expected massive modes on the flux tube.

\subsubsection{Corrections to the LC spectrum for open strings}

\begin{figure}[t]
 \centering
 \begin{minipage}{.48\textwidth}
 \includegraphics[width=\textwidth]{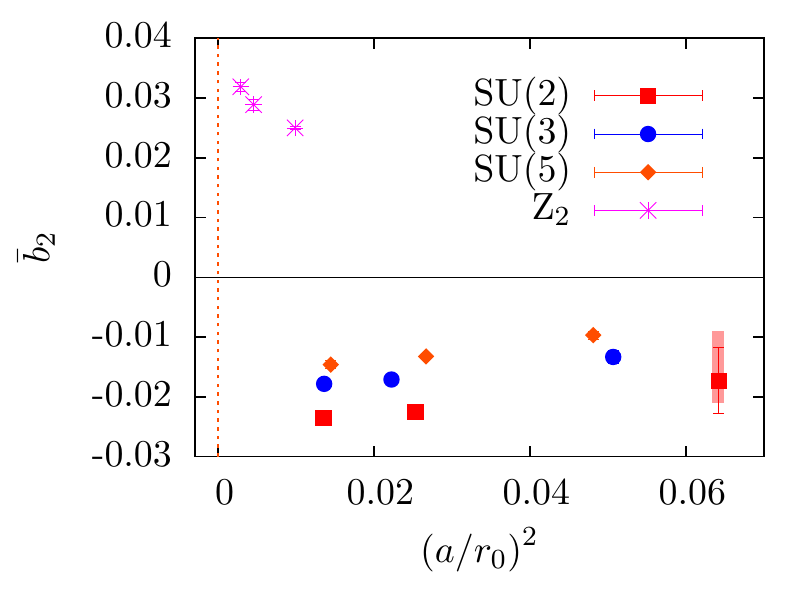}
 \end{minipage}
 \begin{minipage}{.48\textwidth}
 \centering
 \includegraphics[width=.7\textwidth]{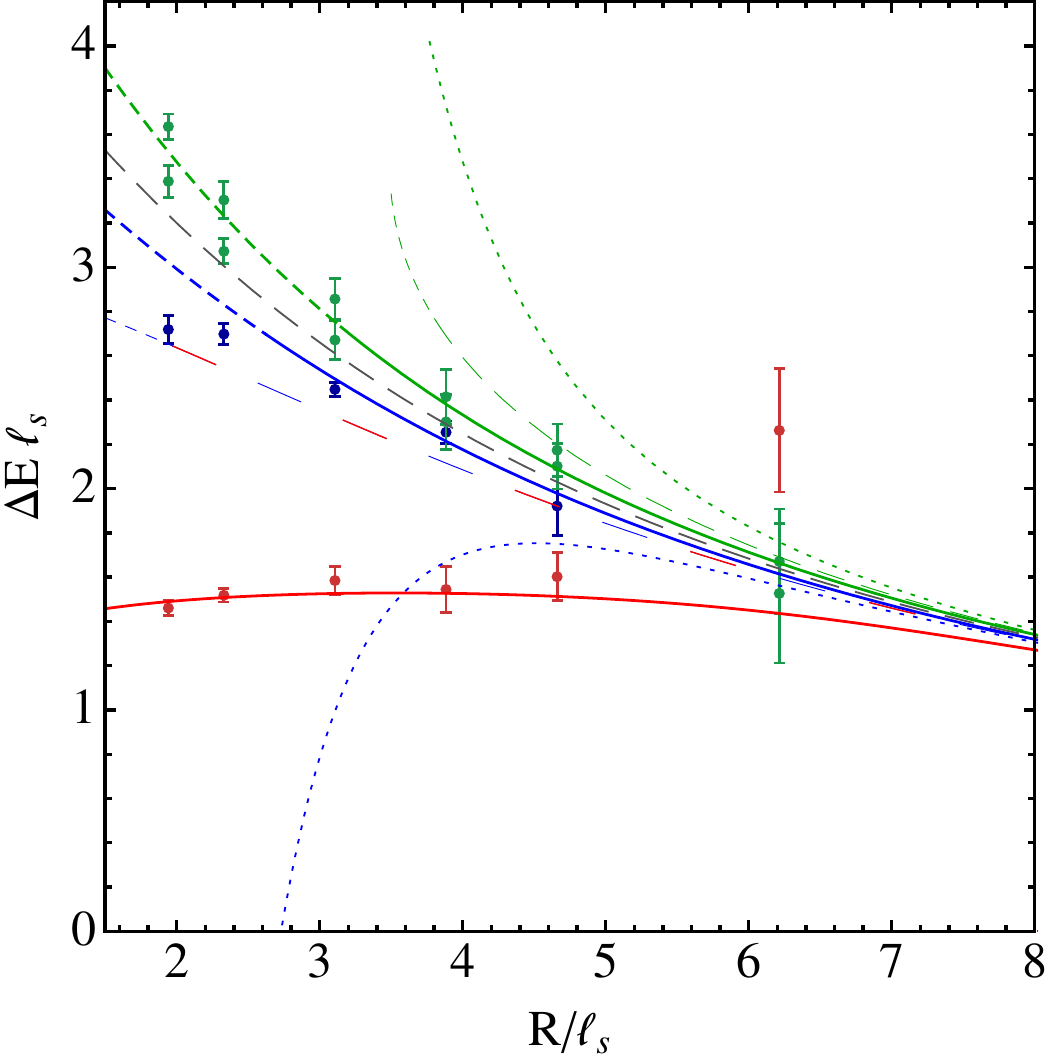}
 \end{minipage}
 \caption{{\bf Left:} Results for the boundary coefficient $\bar{b}_2$ obtained
 from the groundstate energies for $N=2$ and
 3~\cite{Brandt:2013eua,Brandt:unpublished}, preliminary results for
 $N=5$~\cite{Brandt:unpublished}, and results from
 Z$_2$ gauge theory~\cite{Billo:2012da}. The red band is the result for SU(2)
 extracted from excited states~\cite{Brandt:2010bw}. {\bf Right:} Comparison of
 the energy differences to the groundstate of the closed flux tubes in 4D SU(3)
 from~\cite{Athenodorou:2010cs}, with the
 predictions from TBA (solid lines), including the worldsheet axion, taken
 from~\cite{Dubovsky:2013gi}. Scalar, pseudoscalar and spin 2 states are shown
 in blue, red and green, respectively. The dashed lines correspond to the EST
 prediction from TBA without the axion, the dashed gray line is the
 LC prediction and the dotted lines are the LO and NLO orders in the $R^{-1}$
 expansion. Here $\ell_s=\sqrt{\sigma}^{-1}$.}
 \label{fig:b2_plot}
\end{figure}

Concerning the comparison between lattice results and the EST, the open string
channel is advantageous because the first correction not fixed in terms of the
string tension is of $O(R^{-4})$ (compared to $O(R^{-7})$ for closed strings)
and only depends on one free parameter $\bar{b}_2=\sqrt{\si^3}b_2$.\footnote{A
two-loop correction coming from the rigidity term\cite{German:1989vk} also
starts at $O(R^{-4})$ for long strings, but there is no check of compatibility
with Lorentz invariance. This term seems to be present in U(1) gauge
theory,\cite{Caselle:2014eka} and, when present, might contaminate the
extraction of $\bar{b}_2$.} The comparison to the boundary corrections is
conveniently done in the 3D case and the associated coefficient $\bar{b}_2$ has
first been extracted from the excited states in SU(2) gauge
theories~\cite{Brandt:2010bw} and later from the groundstate in
Z$_2$~\cite{Billo:2012da} and in SU($N$) gauge theories for
$N=2,3$~\cite{Brandt:2013eua} ($N=4,5,6$ are in
preparation~\cite{Brandt:unpublished}) at several lattice spacings. Including
the boundary correction, the EST can describe the groundstate data down to about
0.4~fm and is in qualitative agreement with the excited
states~\cite{Brandt:2010bw,Billo:2012da}. In fact, by leaving the exponent of
the correction term free, it can be shown that it takes a value of 4 down to
$R\approx 0.4$~fm~\cite{Brandt:unpublished}. In figure~\ref{fig:b2_plot}, we
collect the results for $\bar{b}_2$ versus the squared lattice spacing. The plot
illustrates the non-universality of $\bar{b}_2$. Interestingly, it appears to
become larger with increasing $N$, possibly tending towards zero for
$N\to\infty$, and is positive for Z$_2$ gauge theory, where, however,
$\bar{b}_2$ does not show a perfect scaling towards the continuum. From the open
string groundstate in 3D U(1) gauge theory, also the rigidity contribution was
extracted\cite{Caselle:2014eka}, thanks to an enhancement of $\al$ towards the
continuum. This is not expected for SU($N$) and indeed, a positive deviation
from the LC spectrum is visible for short strings, while the rigidity term
gives a negative Coulomb-like contribution at small values of
$\sqrt{\si/\al}R$.\cite{German:1989vk}

In 4D, such an analysis is still missing. The
spectrum~\cite{Morningstar:1998da,Juge:2002br,Juge:2004xr} shows a strong
rearrangement of energy levels from short distance~\cite{Bali:2003jq} to long
distance (string) degeneracies. The spectrum graduates in states that show a
fast convergence to the string states and those with an anomalously slow
approach, such as the $\Sigma_u^-$ ($J^{CP}=0^{--}$) state and possibly also the
states ${\Sigma'}_g^-$ and $\Sigma_g^-$ (similar states have been found for
closed flux tubes), which possibly receive contributions from massive
modes~\cite{Kuti:2005xg,Athenodorou:2010cs,Dubovsky:2013gi,Dubovsky:2014fma}.
Note, that the 4D energies
from~\refcite{Morningstar:1998da,Juge:2002br,Juge:2004xr} typically overshoot
the LC predictions in the large $R$ limit, which could be a sign for
contaminations from excited states
(e.g.~\refcite{Majumdar:2002mr,Brandt:2010bw,Athenodorou:2010cs}) and a
continuum limit is still missing.\footnote{Some of these states were
extrapolated to the continuum in~\refcite{Bali:2003jq}, showing small lattice
artefacts.}

\subsubsection{Corrections to the LC spectrum for closed strings}
\label{subsub:lat:closed}

For closed flux tubes, the corrections to the LC spectrum for the groundstate
and excited states in 3D start at $O(R^{-7})$ and are thus harder to detect.
Interestingly, all of them appear to be negative,\footnote{The exception are the
3D SU(2) results, possibly due to finite temperature
effects.\cite{Athenodorou:2011rx,Athenodorou:2016kpd}.} in contrast to the
positive corrections in the open case.\footnote{This is in agreement with the
next $O(R^{-6,7})$ terms observed in the 3D open
case~\cite{Brandt:2010bw,Brandt:unpublished}.} For 4D the correction to excited
states is universal and of $O(R^{-5})$. There is one state with $J^P=0^-$, which
deviates from the string energy levels and thus qualifies as a massive
excitation. Leaving this state aside, an analysis of the corrections to the LC
energy levels, has been performed for 4D~\cite{Athenodorou:2010cs} (mostly for
$N=3$) and 3D~\cite{Athenodorou:2011rx,Athenodorou:2016kpd} (for $N=2,3,4,5,6$
and 8). For the groundstate, the results in general show good agreement with the
EST predictions, leading to an exponent of -7 or -9 (at most -5) for the
corrections to the LC spectrum with $N\geq3$.

For the excited states in 3D a simple $R^{-7}$ correction is basically ruled out
when fitting the data down to
$R\sqrt{\sigma}\approx4.5-5.0$~\cite{Athenodorou:2011rx}. It is likely that the
radius of convergence, $R_C$, of the derivative expansion in the full EST
increases with the excitation level (similar to the one of the LC spectrum). An
analysis with an heuristic resummed correction\cite{Athenodorou:2011rx} yields
$R_C$ of the order of $R_C$ of the LC spectrum. The TBA analysis, which
corresponds to a more convergent expansion, obtains good fits and possibly some
hints of a massive resonance.\cite{Dubovsky:2014fma} For 4D, a simple $R^{-5}$
correction down to $R_C$ of the LC spectrum is not ruled out. Nonetheless, a
similar increase of the deviations to the LC spectrum are seen with increasing
excitation level. For closed flux tubes there are also states with non-zero
longitudinal momentum, $q\neq0$. The lattice results for those states agree with
the above findings and basically consistent with the LC
predictions.~\cite{Athenodorou:2010cs,Athenodorou:2011rx,Athenodorou:2016kpd}

Looking at the state with $J^P=0^-$, both a simple heuristic
analysis,\cite{Athenodorou:2010cs} and the TBA
method,\cite{Dubovsky:2013gi,Dubovsky:2014fma} show agreement of the data
with a massive mode contribution. In particular, the latter explicitly
identifies the worldsheet axion and shows excellent agreement with this state
and the next excited state in this channel. The $0^-$ state with $q=1$ shows the
same behaviour as the associated $q=0$ state. However, the next excited state in
the $q=1$, $0^-$-channel behaves differently, basically following the LC
predictions. It would be interesting to see whether the TBA analysis can explain
the behaviour of this state, too.

\begin{figure}[t]
 \centering
 \begin{minipage}{.48\textwidth}
 \includegraphics[width=.9\textwidth]{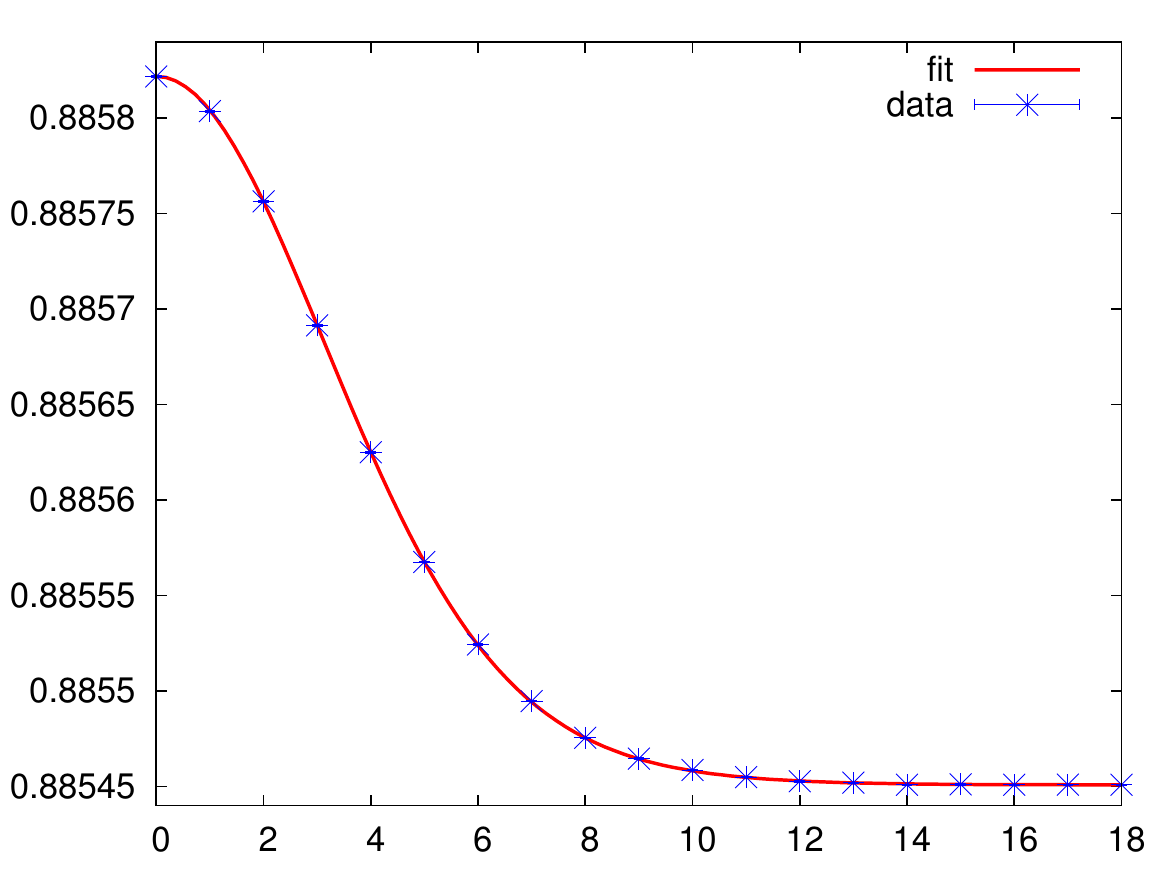}
 \end{minipage}
 \begin{minipage}{.48\textwidth}
 \includegraphics[width=.9\textwidth]{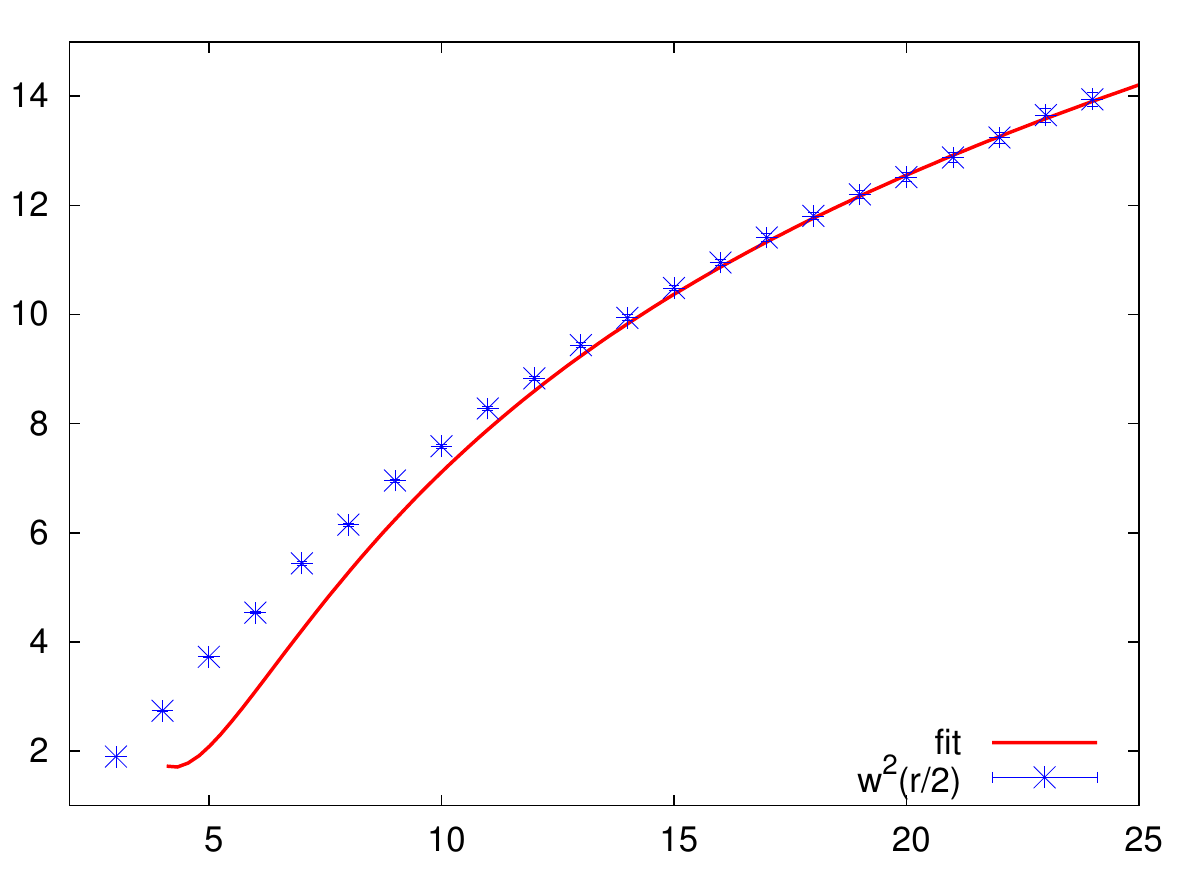}
 \end{minipage}
 \caption{{\bf Left:} Flux tube profile at the midpoint in 3D SU(2)
 from~\citen{Gliozzi:2010zv} for $R\approx1.4$~fm. The $y$ axis shows the
 expectation value of the Polyakov loop correlator in $x_1$ direction
 correlated with a plaquette in $(x_1,x_2)$ direction versus the transverse
 displacement in lattice units ($a\approx0.07$~fm). The solid line is a fit to a
 Gaussian with correction terms (see eq. (20) in~\citen{Gliozzi:2010zv}).
 {\bf Right:} Flux tube width versus its length in
 lattice units, again from~\citen{Gliozzi:2010zv}, including a logarithmic fit.}
 \label{fig:profile}
\end{figure}

\subsection{Width of the flux tube}

Another observable to investigate the range of validity of the EST is the flux
tube profile.\footnote{In the following we will ignore the possible impact of
the considered field strength components on the analysis for simplicity and
brevity.} Studies of the profile have been performed in
SU($N$)~\cite{Fukugita:1983du,Flower:1985gs,Wosiek:1987kx,
Sommer:1987uz,DiGiacomo:1989yp,DiGiacomo:1990hc,Cea:1992vx,Singh:1993jj,
Trottier:1993jv,Matsubara:1993nq,Bali:1994de,Haymaker:1994fm,Cea:1994ed,
Trottier:1995np,Cea:1995zt,Green:1996be,Bali:1997cp,Gubarev:1999yp,
Koma:2003gq,Koma:2003hv,Haymaker:2005py,Chernodub:2005gz,D'Alessandro:2006ug,
Sekido:2007mp,Suzuki:2009xy,Cardaci:2010tb,Gliozzi:2010zv,Cea:2012qw,
Cardoso:2013lla,Cea:2014uja},
U(1)~\cite{Peterson:1984tk,Zach:1997yz,Koma:2003gi,Caselle:2016mqu} and
Z$_2$~\cite{Hasenbusch:1992zz,Caselle:1995fh} gauge theories and confirmed
the formation of a tube-like object. Most studies have compared the profile to
the predictions from the dual-superconductor model only, whose discussion is
beyond the scope of this review.\footnote{To shortly summarize: All of the
studies find good consistency with the exponential decay of the tail of the
profile and newer studies could also describe the inner core with a modified
fitting function~\cite{Cea:2012qw,Cea:2014uja}. Typically the penetration length
is between 0.12 and 0.17~fm and the parameters indicate superconductivity at the
border between types I and II. Note, that all of these studies consider flux
tubes of length smaller than 1.0~fm, which are not yet expected to be well
described by the EST.} Initial
studies~\cite{Sommer:1987uz,DiGiacomo:1989yp,DiGiacomo:1990hc} comparing
to the EST, saw agreement starting from around
$R\approx0.8$~fm~\cite{Bali:1994de}, yet, without being able to clearly
identify the logarithmic broadening with increasing $R$. The latter has first
been confirmed in Z$_2$ gauge theory~\cite{Hasenbusch:1992zz,Caselle:1995fh}.
Similar studies in 3D SU(2)~\cite{Gliozzi:2010zv} and
U(1)~\cite{Caselle:2016mqu} and 4D SU(3) gauge theory~\cite{Cardoso:2013lla}
only became available recently. These studies show the logarithmic broadening of
the string with $R$ and evidence for the Gaussian shape for flux tubes with
$R\gtrsim1.3$~fm. Higher orders could potentially increase the range of
agreement. The results for the profile, including a Gaussian fit with EST
inspired corrections~\cite{Gliozzi:2010zt}, and the width
from~\refcite{Gliozzi:2010zv}, together with a logarithmic fit, are shown in
figure~\ref{fig:profile}. The study from~\refcite{Cardoso:2013lla} focuses on
4D SU(3) and has gone the next step, trying to identify signs of a mixture of
the vortex and string pictures in the profile. They fitted the profile to a
convolution of a Gaussian with an exponential decrease and found a good
description of the data for distances between 0.4 and 1.4~fm. The associated
penetration length remains constant (at 0.22~fm) with $R$, in agreement with the
dual superconductor picture, while the squared width increases logarithmically.
Note, that both studies still lack a continuum extrapolation. In 3D
U(1)~\cite{Caselle:2016mqu}, the width increases logarithmically, but the tail
of the profile does not show a Gaussian shape. It would be interesting to see
whether a combined vortex/string analysis also works for this high precision
case.

\section{Summary and perspectives}

In this review, we have summarized the current knowledge about the theoretical
foundation of EST for confining flux tubes and the associated predictions,
together with comparisons to simulations in lattice QCD. In sec.
\ref{sec:action} and \ref{sec:predict}, we have accumulated the new theoretical
insights of the last few years in a homogeneous presentation of the EST. As
shown in sec. \ref{sec:test}, on one hand predicted deviations from the LC
spectrum are in good agreement with lattice results. On the other hand, the TBA
approach has allowed a solid interpretation of anomalous data in 4D in terms of
a massive pseudoscalar mode. For the width of the flux tube, simulations show
good agreement with the EST predictions starting from around 1 to 1.3~fm.

Despite the good agreement, there are several open questions. To begin with, it
would be interesting to improve on the precision for the excited states in 3D
and to reliably perform the analyses concerning corrections to the spectrum for
4D. In particular, the contribution of the axion to the open string spectrum has
not been computed and a general understanding of other possible massive modes is
lacking. Concerning the profile, the competition between the exponential and the
Gaussian tail highlights the need for more theoretical control over the simple
idea of the flux tube as a vortex with stringy fluctuations on
top.~\cite{Cardoso:2013lla}. One more ripple in the consolidated understanding
of the EST is provided by the contribution of the rigidity term to the
observables, which accounts for some features of the spectrum in U(1) gauge
theory and may appear in other gauge theories, too. It is crucial to confirm
these effects via an explicitly Lorentz invariant computation. On a more
speculative level, the availability of first principle computations of the
effective action in holographic
setup\cite{Aharony:2009gg,Kol:2010fq,Vyas:2010wg,Vyas:2012dg,Giataganas:2015yaa}
fuels the hope that one might, in turn, extract from the data useful information
about the holographic dual of Yang-Mills theories. Finally, it would be
interesting to gain more insight about modifications coming from the finite
masses of `static' quarks and the presence of sea quarks.

There are some issues related to flux tubes and the EST that we could not
discuss due to length constraints. This concerns signatures of the EST at finite
temperature~\cite{Giudice:2009di,Caselle:2011vk,Caselle:2015tza} and the
behaviour of the width in this
regime~\cite{Gliozzi:2010jh,Bakry:2010zt,Caselle:2012rp,Cea:2015wjd}. One can
also study flux tubes in different
representations~\cite{Pepe:2009in,Athenodorou:2013ioa,Athenodorou:2016kpd}, so
called $k$-strings, baryonic
boundstates~\cite{deForcrand:2005vv,Pfeuffer:2008mz,Bakry:2014gea} or
the interface free energy in 3D Z2 gauge
theory~\cite{Caselle:2006dv,Caselle:2007yc}. These studies typically observe
rather good agreement with the EST and massive modes have also been observed for
$k$-strings~\cite{Athenodorou:2013ioa,Athenodorou:2016kpd}. There are also other
aspects of the potential which have not been discussed
(see~\refcite{Greensite:2003bk,Bali:2000gf}, for instance).

\subsection*{Acknowledgements}

We acknowledge very enlightening discussions and correspondence with A.
Athenodorou, M. Caselle, F. Cuteri, R. Flauger, D. Gaiotto, P. Majumdar and M.
Panero. Research at Perimeter Institute is supported by the Government of Canada
through Industry Canada and by the Province of Ontario through the Ministry of
Research \& Innovation. BB receives funding by the DFG via SFB/TRR 55 and the
Emmy Noether Programme EN 1064/2-1.

\end{document}